\documentclass[a4paper,12pt]{article}

\usepackage[left=2.5cm,right=2.5cm,top=2.5cm,bottom=2.5cm]{geometry}
\usepackage{graphicx,amssymb,amsmath,amsthm,setspace,subcaption,cite,authblk}

\usepackage[colorlinks=true,bookmarks=true]{hyperref}
\usepackage{breakurl} 

\usepackage{mathtools}

\DeclarePairedDelimiterX\braket[2]{\langle}{\rangle}{#1 \delimsize\vert #2}

\hypersetup{citecolor=blue}
\newcommand{\dif}{\mathrm{d}}
\newcommand{\Eqref}[1]{(\ref{#1})}
\newcommand{\half}{\frac{1}{2}}

\newcommand{\brac}[1]{\left(#1 \right)}
\newcommand{\sbrac}[1]{\left[#1\right]}

\newcommand{\im}{\mathrm{i}}
 
\begin{document}

\title{Null geodesics in the C-metric with cosmological constant}

\author{Yen-Kheng Lim\footnote{Email: yenkheng.lim@xmu.edu.my}}

\affil{\normalsize{\textit{Department of Physics, Xiamen University Malaysia, 43900 Sepang, Malaysia}}}

\date{\normalsize{\today}}
\maketitle
 
\renewcommand\Authands{ and }
\begin{abstract}
 In this paper we study the motion of photons or massless particles in the C-metric with cosmological constant. The Hamilton--Jacobi equations are known to be completely separable, giving a Carter-like quantity $Q$ which is a constant of motion. All possible trajectories are classified according to a two-dimensional parameter space representing the particle's angular momentum and energy scaled in units of $Q$. Exact solutions are given in the C-metric coordinates in terms of Jacobi elliptic functions. Using the exact solutions, we find examples of periodic orbits on the photon surface.
\end{abstract}


\section{Introduction} \label{intro}

The study of null geodesics is a useful tool in revealing the features of strong gravity in various spacetimes. This is perhaps becoming increasingly so in recent years as direct imaging and detection of gravitational waves have become observational realities. On the theoretical side, studying the orbits of light may provide clues to solutions of wave equations and massless fields in the background of a certain spacetime. In this paper, we study the classical motion of light propagating in the C-metric in Einstein gravity with a cosmological constant.

The Ricci-flat C-metric \cite{Levi-Civita:1918,Weyl1919}, in its maximally-extended description, describes two causally-disconnected black holes uniformly accelerating apart. In this paper we will confine ourselves in a particular Lorentzian patch where only one accelerating black hole is observed. The cause of this acceleration is either a cosmic strut providing positive pressure or cosmic string providing positive tension. A review of the geometrical properties and global structure for the Ricci-flat C-metric can be found in \cite{Griffiths:2006tk,Griffiths:2009dfa}. The quasi-normal modes and stability properties of the C-metric have been studied by Destounis et al. in Refs.~\cite{Destounis:2020pjk,Destounis:2020yav}. The C-metric solution can be straightforwardly generalised to include a positive or negative cosmological constants. These describe accelerating black holes in a de Sitter (dS) or anti-de Sitter (AdS) backgrounds, respectively. The structure and properties the (A)dS C-metrics have been studied in Refs.~\cite{Podolsky:2000pp,Podolsky:2002nk,Dias:2002mi,Dias:2003xp,Krtous:2003tc,Podolsky:2003gm,Krtous:2005ej}. In particular, the solutions in AdS backgrounds may be of interest in the studies of the AdS/CFT correspondence. For instance, the minimal surfaces in the AdS C-metric studied in \cite{Xu:2017nut} might be relevant in evaluating the holographic entanglement entropy of the dual theory \cite{Tavakoli:2018xnh}.

In the (A)dS C-metric, the equations of motion for the null geodesics are completely integrable. In this paper we derive them by separating the Hamilton--Jacobi equations. Previously, the study of geodesics in the C-metric was done by \cite{Pravda:2000zm}, and in \cite{Chamblin:2000mn} for the Anti-de Sitter C-metric. The general equations for null and time-like geodesics in the Ricci-flat case was studied the author's earlier paper \cite{Lim:2014qra}, as well as by Bini et al. in \cite{Bini:2014iga}. In Ref.~\cite{Alawadi:2020qdz} Alawadi, Batic, and Nowakowski studied circular photon orbits in the Ricci-flat C-metric, where it was rigorously shown that such photon rings are unstable. In Ref.~\cite{Frost:2020zcy}, Frost and Perlick studied the equations of motion of null geodesics and gravitational lensing, also in the Ricci-flat C-metric. Most recently, Zhang and Jiang studied null geodesics and black-hole shadows in the rotating generalisation of the Ricci-flat C-metric \cite{Zhang:2020xub}. Black hole shadows and photon spheres have been studied in various other spacetimes, e.g., in \cite{Mishra:2019trb}.

In this paper, we shall consider a more generalised setting where a non-zero cosmological constant 
is present, and attempt to place a higher emphasis on aspects not yet covered by these recent papers. We attempt a classification of possible orbits by in terms of a two-dimensional parameter space related to the photon's angular momentum and energy scaled in units of the Carter-type separation constant $Q$. The parameter space will be organised based on the allowed domains of existence of the geodesic. The analysis and exact solutions derived in the present paper will be done in the C-metric $(x,y)$-coordinates. Additionally, we also obtain examples of periodic orbits where photons move along closed trajectories on the photon surface.


The rest of the paper is organised as follows. In Sec.~\ref{EOM} we review the essential features of the C-metric spacetime and derive the equations of motion for photon orbits. In Sec.~\ref{ranges} we study the parameter space of the orbits and classify the different possible types in a two-dimensional parameter space. With the separated Hamilton--Jacobi equations, we obtain exact solutions in Sec.~\ref{analytical}. Using the analytical solutions, we obtain some examples of periodic orbits in Sec.~\ref{closed}. Conclusions and closing remarks are given in Sec.~\ref{conclusion}.

\section{The C-metric and geodesic equations} \label{EOM}

We will take our C-metric to be in the following form 
\begin{subequations} \label{Cmetric}
\begin{align}
 \dif s^2&=\frac{1}{A^2(x-y)^2}\brac{-F(y)\dif t^2+\frac{\dif y^2}{F(y)}+\frac{\dif x^2}{G(x)}+G(x)\dif\phi^2}, \\
 F(y)&=\frac{1}{\ell^2A^2}-\brac{1-y^2}\brac{1+2mAy},\quad G(x)=\brac{1-x^2}\brac{1+2mAx},
\end{align}
\end{subequations}
where $m\geq0$ and $A>0$ are respectively the mass and acceleration parameters of the C-metric, and $\ell$ is the (anti)-de Sitter curvature parameter related to the cosmological constant $\Lambda$ by $\ell^{-2}=-\frac{\Lambda}{3}$. For this value, the metric \Eqref{Cmetric} solves the Einstein equation $R_{\mu\nu}=\Lambda g_{\mu\nu}$. Since we define $\ell^2$ to have a sign opposite to the cosmological constant, we henceforth refer to the case $\ell^2>0$ as the \emph{anti-de Sitter (AdS) C-metric} and $\ell^2<0$ the \emph{de Sitter (dS) C-metric}. The Ricci-flat C-metric is the limit $\ell^2\rightarrow\pm\infty$, for which we recover precisely the fully factorised form of Hong and Teo \cite{Hong:2003gx}. An alternative form of the (A)dS C-metric was proposed in \cite{Chen:2015vma}, where $F$ and $G$ are both factorised. For the studies its geodesic equations, we find it more convenient to use the form in \Eqref{Cmetric} where the expressions for certain parameters appear more compact. Besides these, the forms of $F$ and $G$ where they are unfactorised was used in, for instance, Refs.~\cite{Podolsky:2003gm,Krtous:2003tc} in the study of gravitational radiation, and in Refs.~\cite{Podolsky:2000pp,Dias:2002mi,Dias:2003xp,Krtous:2005ej} in obtaining the causal structure of the (A)dS C-metrics.

In this form, $G$ remains factorised with its roots easily readable as $x=\pm1$, $-\frac{1}{2mA}$, but $F$ is not in factorised form due to the presence of the term $1/\ell^2A^2$. Let us denote the three possible roots of $F$ by $\{a,b,c \}$ with the ordering $a\leq b\leq c$ whenever they are real. The three roots are real and distinct if 
\begin{align}
 \ell^2>\frac{54m^2A^2\Omega_+}{2A^2(1-4m^2A^2)^2}\quad\mbox{ or }\quad\ell^2<\frac{54m^2A^2\Omega_-}{2A^2(1-4m^2A^2)^2}, \label{Froots}
\end{align}
where 
\begin{align}
 \Omega_\pm=\frac{1-36m^2A^2\pm\brac{1+12m^2A^2}^{3/2}}{54m^2A^2} \label{Omega_def}
\end{align}
For $0<mA<\half$, we always have $\Omega_+>0$ and $\Omega_-<0$. Therefore the domain $\ell^2>\frac{54m^2A^2\Omega_+}{2A^2(1-4m^2A^2)^2}$ lies in the AdS case and $\ell^2<\frac{54m^2A^2\Omega_-}{2A^2(1-4m^2A^2)^2}$ lies in the dS case.
If $\ell^2$ approaches $\frac{54m^2A^2\Omega_-}{2A^2(1-4m^2A^2)^2}$ from below, $a$ approaches $b$, and if $\ell^2$ approaches $\frac{54m^2A^2\Omega_+}{2A^2(1-4m^2A^2)^2}$ from above, $b$ approaches $c$. Evaluating the derivative of $F$ at $y=0$, we find $F'(0)=-2mA$, which is negative. The constant term of $F$ is $1-1/\ell^2A^2$. Therefore we conclude that $b\leq 0$ if $\ell^2A^2\geq 1$, and $b>0$ if $\ell^2A^2<1$. In Ref.~\cite{Dias:2002mi}, it was shown that the former case corresponds to a single accelerated black hole in AdS, and the latter describes two accelerating black holes in AdS.

We are interested in parameter ranges where $G\geq0$ and $F\geq0$ for non-empty domains 
\begin{align}
 -1\leq x\leq 1,\quad a\leq y\leq b, \label{Lorentzian}
\end{align}
where the metric will be static and Lorentzian with signature $(-,+,+,+)$. The coordinate $x$ is interpreted as a polar-type coordinate where, for convenience of exposition, we will refer to the roots $G(\pm1)=0$ as the `north' and `south' poles, respectively. We will sometimes refer to $x=0$ as the `equator'. The coordinate $y$ can be interpreted as a radial-type coordinate distance from the black hole. Therefore, observers in the domain \Eqref{Lorentzian} will see an accelerating black hole with a horizon of spherical topology located at $y=a$, as well as an acceleration horizon $y=b$. To ensure that such domains exist, we will restrict our attention to the parameters satisfying
\begin{align}
 2mA<1,\quad \ell^2\left\{\begin{array}{cc}
                           >\frac{54m^2A^2\Omega_+}{2A^2(1-4m^2A^2)^2} & \mbox{AdS case},\\
                           <\frac{54m^2A^2\Omega_-}{2A^2(1-4m^2A^2)^2} & \mbox{dS case}.
                          \end{array}
 \right.
\end{align}
The Ricci-flat C-metric is recovered if the limit $\ell^2\rightarrow+\infty$ is taken from the AdS side, or if $\ell^2\rightarrow-\infty$ is taken from the dS side. In either limit, $a$ and $b$ become $-1$ and $+1$, respectively.

The spacetime carries an unavoidable conical singularity on either of the $x=-1$ or $x=1$ half-axes. For instance we can choose to remove the singularity at $x=1$ by fixing the periodicity of the azimuthal angle to be 
\begin{align}
 \phi=\phi+\Delta\phi,\quad \Delta\phi=\frac{2\pi}{1+2mA}.
\end{align}
Equivalently one can define a new coordinate via $\phi=\frac{\varphi}{1+2mA}$ such that $\varphi\in[0,2\pi]$ runs through the usual periodicity. With the removal of the $x=1$ singularity the C-metric is interpreted as the black hole being pulled `southwards' ($x=-1$) by a cosmic string. However, for the purposes of this paper, we have no particular reason to favour the removal of a cosmic strut over the cosmic string, or vice versa. So when numerical examples are considered, we shall simply take $\Delta\phi=2\pi$.

The trajectory of particles in the C-metric shall be described by a parametrised curve $q^\mu(\tau)=\brac{t(\tau),y(\tau),x(\tau),\phi(\tau)}$ where $\tau$ is an affine parameter. In this paper we shall use the terms \emph{photons}, \emph{light}, and \emph{massless particles} interchangeably to mean particles that follow geodesics where their tangent vectors are null, $g_{\mu\nu}\dot{q}^\mu\dot{q}^\nu=0$. (Over-dots denote derivatives with respect to $\tau$.) The equations of motion can be derived starting from the Lagrangian $\mathcal{L}=\half g_{\mu\nu}\dot{q}^\mu\dot{q}^\nu$, from which the Euler--Lagrange equations $\frac{\dif}{\dif\tau}\frac{\partial\mathcal{L}}{\partial\dot{q}^\mu}=\frac{\partial\mathcal{L}}{\partial q^\mu}$ provides the equations of motion of the particle.

For the C-metric \Eqref{Cmetric}, the Lagrangian is explicitly
\begin{align}
 \mathcal{L}=\frac{1}{2A^2(x-y)^2}\brac{-F\dot{t}^2+\frac{\dot{y}^2}{F}+\frac{\dot{x}^2}{G}+G\dot{\phi}^2}. \label{Lagrangian}
\end{align}
The momenta $p_\mu=\frac{\partial\mathcal{L}}{\partial\dot{q}^\mu}$ conjugate to the four coordinates are
\begin{subequations}\label{momenta}
\begin{align}
 p_t&=-\frac{F}{A^2(x-y)^2}\dot{t}\equiv-E,\label{def_E}\\
 p_\phi&=\frac{G}{A^2(x-y)^2}\dot{\phi}\equiv\Phi, \label{def_Phi}\\
 p_y&=\frac{1}{A^2(x-y)^2}\frac{\dot{y}}{F},\\
 p_x&=\frac{1}{A^2(x-y)^2}\frac{\dot{x}}{G}.
\end{align}
\end{subequations}
Since $\partial_t$ and $\partial_\phi$ are Killing vectors of the spacetime, the momenta along these directions are conserved along the geodesics. We will regard these constants as the energy and angular momentum of the particle denoted by $E$ and $\Phi$ respectively. 

Using Eqs.~\Eqref{def_E} and \Eqref{def_Phi} to eliminate $\dot{t}$ and $\dot{\phi}$ in favour of $E$ and $\Phi$, the null condition reduces to  
\begin{align}
 V+\frac{\dot{y}^2}{F}+\frac{\dot{x}^2}{G}=0,\label{constraint}
\end{align}
where $V$ is our effective potential given by\footnote{This definition of effective potential is slightly different from the one used in \cite{Lim:2014qra}.} 
\begin{align}
 V=A^4(x-y)^4\brac{\frac{\Phi^2}{G}-\frac{E^2}{F}}. \label{V_eff}
\end{align}
Since $F$ and $G$ are positive in the static Lorentzian patch \Eqref{Lorentzian}, Eq.~\Eqref{constraint} constrains the photons to domains of $(x,y)$ where $V\leq0$. In the author's previous work \cite{Lim:2014qra}, this was used to visualise the regions accessible to time-like particles and photons. However, as we will see below, to the separability of the equations for photons will provide a stronger restriction of their domains of existence.

Applying the Euler--Lagrange equations to $x$ and $y$, we have \cite{Lim:2014qra} 
\begin{subequations}\label{ddots}
\begin{align}
 \ddot{x}=&\brac{\frac{G'}{2G}+\frac{1}{x-y}}\dot{x}^2-\frac{G\dot{y}^2}{(x-y)F}-\frac{2\dot{x}\dot{y}}{x-y}+\nonumber\\
          &\hspace{1.5cm}+A^4(x-y)^3G\sbrac{\frac{E^2}{F}+\brac{\frac{(x-y)G'}{2G}-1}\frac{\Phi^2}{G}},\label{xddoteqn}\\
 \ddot{y}=&\brac{\frac{F'}{2F}-\frac{1}{x-y}}\dot{y}^2+\frac{F\dot{x}^2}{(x-y)G}+\frac{2\dot{x}\dot{y}}{x-y}\nonumber\\
          &\hspace{1.5cm}-A^4(x-y)^3F\sbrac{\brac{\frac{(x-y)F'}{2F}+1}\frac{E^2}{F}-\frac{\Phi^2}{G}}, \label{yddoteqn}
\end{align}
\end{subequations}
where we have denoted $G'=\frac{\dif G}{\dif x}$ and $F'=\frac{\dif F}{\dif y}$. Solutions to Eqs.~\Eqref{ddots} have already been studied in the author's earlier paper \cite{Lim:2014qra} in the case of zero cosmological constant. Here numerical solutions to \Eqref{ddots} for zero and non-zero cosmological constants are used as an independent consistency check against the analytical results to be derived in Secs.~\ref{ranges} and \ref{analytical}.

Before separating the equations of null geodesics in the C-metric \Eqref{Cmetric}, we first point out that the general family of Pleba\'{n}ski--Demia\'{n}ski spacetimes admits a hidden symmetry described by a Killing--Yano 2-form \cite{Kubiznak:2007kh} rendering the equations for null geodesics separable. Since the C-metric is a member of the Pleba\'{n}ski--Demia\'{n}ski family, it inherits the separability of the Hamilton--Jacobi equations. The separation procedure was done, e.g., in \cite{Grenzebach:2015oea,Sharif:2015pbw,Zhang:2020xub} which includes the rotating version of the spacetime. In the case of the non-rotating AdS C-metric, the separation of the geodesic equations was also performed in \cite{Podolsky:2003gm}, where the separation constant plays a role in the study of gravitational radiation of the spacetime. 

Here let us briefly review the procedure for the present form of the metric \Eqref{Cmetric}. To start, we seek the Hamilton--Jacobi equation $\mathcal{H}\brac{q,\frac{\partial S}{\partial q}}+\frac{\partial S}{\partial\tau}=0$ where $\mathcal{H}$ is the Hamiltonian is obtained by a Legendre transform of \Eqref{Lagrangian}. Explicitly, it is
\begin{align}
 \mathcal{H}\brac{q,p}=\half A^2(x-y)^2\brac{-\frac{p_t^2}{F}+Fp_y^2+Gp_x^2+\frac{p_\phi^2}{G}}.
\end{align}
The Hamilton--Jacobi equation is then 
\begin{align}
 \half A^2(x-y)^2\sbrac{-\frac{1}{F}\brac{\frac{\partial S}{\partial t}}^2+F\brac{\frac{\partial S}{\partial y}}^2+G\brac{\frac{\partial S}{\partial x}}^2+\frac{1}{G}\brac{\frac{\partial S}{\partial\phi}}^2}+\frac{\partial S}{\partial\tau}=0.\label{HJE}
\end{align}
We take the ansatz for Hamilton's principal function $S$ to have the separable form 
\begin{align}
 S=-Et+\Phi\phi+S_x(x)+S_y(y).
\end{align}
Substitution of the ansatz into \Eqref{HJE} leads us to a Carter-like separation constant $Q$ where the equation separates to the pair 
\begin{align}
 F\brac{\frac{\dif S_y}{\dif y}}^2=-\frac{Q}{F}+\frac{E^2}{F^2},\quad G\brac{\frac{\dif S_x}{\dif x}}^2=\frac{Q}{G}-\frac{\Phi^2}{G^2}.
\end{align}
Then, with the relations $p_\mu=\frac{\partial S}{\partial q^\mu}$, along with \Eqref{momenta}, we obtain the set of first-order differential equations
\begin{subequations}\label{EOM1o}
\begin{align}
 \frac{\dot{t}}{A^2(x-y)^2}&=\frac{E}{F},\label{tdot}\\
 \frac{\dot{\phi}}{A^2(x-y)^2}&=\frac{\Phi}{G},\label{phidot}\\
 \frac{\dot{x}}{A^2(x-y)^2}&=\pm\sqrt{X(x)},\label{xdot}\\
 \frac{\dot{y}}{A^2(x-y)^2}&=\pm\sqrt{Y(y)},\label{ydot} 
\end{align}
\end{subequations}
where $X$ and $Y$ are third-degree polynomials given by
\begin{align}
 X(x)&=QG-\Phi^2, \label{X_def} \\ 
 Y(y)&=-QF+E^2. \label{Y_def}
\end{align}
Indeed, Eqs.~\Eqref{tdot} and \Eqref{phidot} are reproductions of \Eqref{def_E} and \Eqref{def_Phi}; and eliminating $Q$ between \Eqref{xdot} and \Eqref{ydot} recovers \Eqref{constraint}. 

\section{Parameter and coordinate ranges of the geodesics} \label{ranges}

By investigating the root structure of $X$ and $Y$, we will be able to characterise the different possible types of orbits. From Eqs.~\Eqref{xdot} and \Eqref{ydot}, the geodesics are restricted to domains where $X\geq0$ and $Y\geq0$. The boundaries of these domains are the roots of $X$ and $Y$. Notice that these two polynomials are obtained by $G$ and $F$ by multiplication and shifts by constants. Therefore the root structures of $X$ and $Y$ are closely related to those of $G$ and $F$. In particular, the quantities $\Omega_\pm$ defined in Eq.~\Eqref{Omega_def} continue to play a role here.

\subsection{The root structure of \texorpdfstring{$X$}{X}}

As mentioned above, the geodesics can only exist in the domains of $x$ where $X\geq0$. This condition immediately requires $Q\geq0$, and that $Q=0$ is only possible for geodesics of zero angular momentum. 

\begin{figure}
 \begin{center}
  \includegraphics{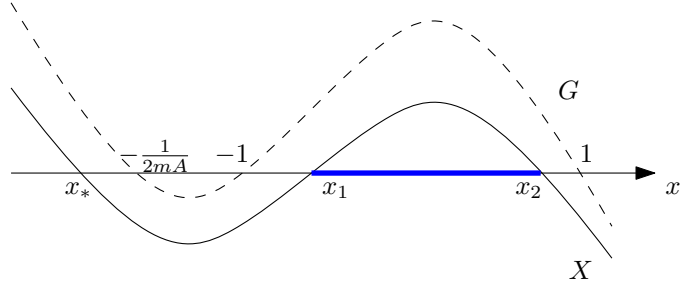}
  \caption{The graphs of $X$ (solid curve) and $G$ (dashed curve) as functions of $x$. The allowed domains for null geodesics correspond to $X\geq0$ and $G>0$. The physically relevant domains satisfying this condition are marked blue.}
  \label{fig_GXgraph}
 \end{center}

\end{figure}

Let us denote the roots by $\{x_*,x_1,x_2\}$ with the ordering $x_*\leq x_1\leq x_2$ whenever they are real. Assuming for the moment that this is the case, we see from Eq.~\Eqref{X_def} that $X$ is obtained from $G$ by multiplication with $Q$, then shifted down by $\Phi^2$ units, as sketched in Fig.~\ref{fig_GXgraph}. From the figure we infer that the relative positions of the roots of $G$ and $X$ are 
\begin{align}
 x_*\leq\gamma<-1\leq x_1<x_2\leq 1,
\end{align}
and that geodesics exist in the domain 
\begin{align}
 x_1\leq x\leq x_2,
\end{align}
for which $X\geq0$, this domain is marked blue in Fig.~\ref{fig_GXgraph}. The other domain $x\leq x_*$ is outside \Eqref{Lorentzian} and is therefore irrelevant.

The reality of the roots are determined by the discriminant
\begin{align}
 \Delta(X)&=4Q^2\sbrac{-27m^2A^2\Phi^4-\brac{1-36m^2A^2}Q\Phi^2+\brac{1-4m^2A^2}Q^2}.
\end{align}
The function $X$ will have real and distinct roots if $\Delta(X)>0$, or  
\begin{align}
 -\Omega_+Q<\Phi^2<-\Omega_-Q,
\end{align}
where $\Omega_\pm$ is as defined in Eq.~\Eqref{Omega_def}. If $\Phi^2>-\Omega_-Q$, $x_1$ and $x_2$ become a complex conjugate pair, while if $\Phi^2<-\Omega_+Q$, the complex conjugate pair is $x_*$ and $x_1$. As we have previously noted, $\Omega_+$ is positive for $0<mA<\half$. Therefore this latter case is physically irrelevant. 

With the Descartes rule of signs, we can further pin down the specific regions occupied by the roots. Using this rule, we see that if $Q-\Phi^2>0$, $X$ has one positive root, and if $Q-\Phi^2>0$, $X$ has two positive roots. Furthermore, since $-\Omega_-$ is always greater than $1$ for $0<mA<\half$, then $Q-\Phi^2$ crosses from negative to positive before reaching $\Phi^2=-\Omega_-Q$, where the discriminant vanishes. In other words, $x_1$ crosses from negative to positive before coalescing with $x_1$ in the positive domain.

From these considerations, we may organise the $x$-motion of the geodesics into five cases:

\textbf{Case A, $\Phi^2=0$.} These are geodesics with zero angular momentum in the $z$-component. In this case,  $x_1=-1$ and $x_2=1$ and the photon can access the full Lorentzian domain $-1<x<1$. By Eq.~\Eqref{phidot}, $\phi$ is constant for these geodesics. These are polar orbits that intersects with the north or south axis of the C-metric (and possibly colliding with the cosmic string/strut).

\textbf{Case B, $0<\Phi^2<Q$.} In this case the motion is restricted to $x_1\leq x\leq x_2$ which is a proper subset of the Lorentzian region \Eqref{Lorentzian}. The root $x_1$ is negative while $x_2$ is positive, so this domain contains the `equatorial plane' $x=0$. The distinguishing feature of this case is that the geodesics oscillates in the polar direction between $x_1$ and $x_2$, and crosses the `equatorial plane' $x=0$ in between. Increasing the angular momentum narrows the range between $x_1$ and $x_2$. As $\Phi^2$ is increased towards $Q$, the root $x_1$ approaches zero.

\textbf{Case C, $\Phi^2=Q$.} In this case, the constant term of $X$ vanishes and $0$ is one of its roots. By continuity from the previous case, we identify this root to be $x_1=0$. The other two roots can be written explicitly as 
\begin{align}
  x_*=\frac{-1-\sqrt{1+16m^2A^2}}{4mA},\quad x_2=\frac{-1+\sqrt{1+16m^2A^2}}{4mA}.
\end{align}
In this case, the photon's domain of polar oscillation is bounded at one side by the `equatorial plane' as measured in coordinates $x=\cos\theta=0$, and the `southern hemisphere' is no longer accessible to the photons. 

\textbf{Case D, $Q<\Phi^2<-\Omega_-Q$.} By the Descartes rule of signs, there are now at least two positive roots. This is by continuity from the previous cases as $x_1$ crosses the past the point $x=0$ and now becomes a positive root. Therefore the roots in this case are ordered by 
\begin{align}
 x_*<0<x_1< x_2<1.
\end{align}
In this case, the domain of polar oscillation is now strictly to the north of the 'equatorial plane'. Further increasing $\Phi^2$ towards $Q\Omega_-$ continues to narrow the domain as $x_1$ approaches $x_2$.

\textbf{Case E, $\Phi^2=-\Omega_-Q$.} Here the discriminant $\Delta(X)$ vanishes and the two roots $x_1$ and $x_2$ coalesce into a degenerate root given by
 \begin{align}
  x_1=x_2\equiv x_{\mathrm{pc}}=\frac{-\brac{1-\sqrt{1+12m^2A^2}}}{6mA}.
 \end{align}
Since this is a degenerate root, we have $X(x_{\mathrm{pr}})=X'(x_{\mathrm{pr}})=0$. By Eq.~\Eqref{xdot}, we see that this trajectory corresponds to a constant $x=x_{\mathrm{pc}}>0$ fixed at a polar angle located north of the `equator'. This is the case referred to by Ref.~\cite{Grenzebach:2015oea} as the \emph{photon cone}. The other root is $x_*=\frac{-\brac{1+\sqrt{1+12m^2A^2}}}{6mA}$, which is still located beyond the physically relevant range of interest.

\begin{figure}
 \begin{center}
  \includegraphics{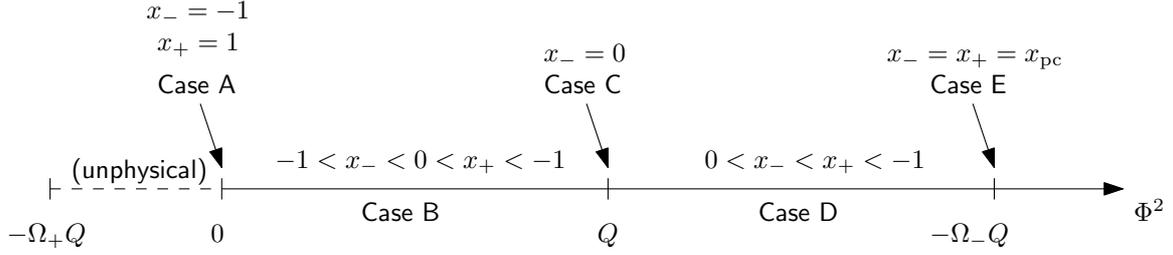}
  \caption{The root structure of $X$, for various values of $\Phi^2$. The quantity $\Omega_\pm$ is as defined in Eq.~\Eqref{Omega_def}.}
  \label{fig_Xrange}
 \end{center}

\end{figure}

For values of $\Phi^2>-\Omega_-Q$ the two roots $x_1$ and $x_2$ become complex and $X$ is negative in the domain $-1\leq x\leq 1$. No geodesics can exist in this case. The cases described above can be summarised in Fig.~\ref{fig_Xrange}, showing the root structure of $X$ as $\Phi^2$ varies from $0$ to $-\Omega_-Q$.  

\subsection{The root structure of \texorpdfstring{$Y$}{Y}}

Let us denote the roots of $Y$ by $\{y_1,y_2,y_* \}$ with the ordering $y_1\leq y_2\leq y_*$ whenever they are real. The constant term of $Y$ is always positive. The Descartes rule of sign tells us that there is only one positive root, which would be the largest, $y_*$.

\begin{figure}
 \begin{center}
  \includegraphics{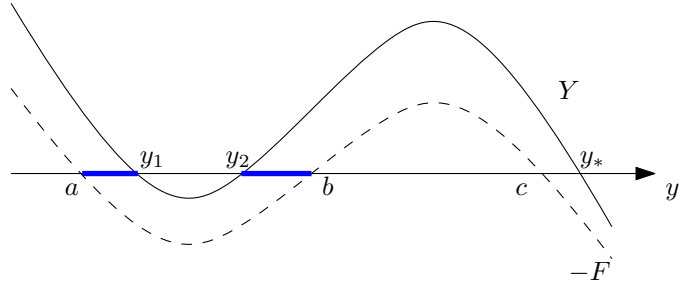}
  \caption{(Colour online) The graphs of $Y$ (solid curve) and $-F$ (dashed curve) as functions of $y$. The allowed domains for null geodesics correspond to $X\geq0$ and $-F<0$. The physically relevant domains satisfying this condition are marked blue.}
  \label{fig_FXgraph}
 \end{center}

\end{figure}

We note that Eq.~\Eqref{ydot} requires $E^2$ to be non-zero. So in the following we consider $E^2>0$. Assuming for the moment that all roots of $F$ and $Y$ are real, we notice that $Y$ is obtained from $-F$ by multiplication of $Q$, then shifted upwards by $E^2$ units. The existence of geodesics requires $Y\geq0$. Recalling that the domain for our static Lorentzian patch is where $F>0$, the relevant domain for null geodesics are marked blue in Fig.~\ref{fig_FXgraph}. From this graph we infer the roots are ordered by
\begin{align}
 a\leq y_1\leq y_2\leq b\leq c\leq y_*
\end{align}

From Fig.~\ref{fig_FXgraph}, we can see that as $E^2$ increases, the graph of $Y$ is raised further up and eventually $y_1$ coalesces with $y_2$, becoming complex. The precise value of $E^2$ where this happens can be determined from the discriminant of $Y$,  
\begin{align}
 \Delta(Y)&=4Q^2\bigg[-27m^2A^2E^4+\brac{1-36m^2A^2+54\frac{m^2}{\ell^2}}QE^2\nonumber\\
   &\hspace{2cm}+\brac{\brac{1-4m^2A^2}^2-\frac{1-36m^2A^2}{A^2\ell^2}-\frac{27m^2}{A^2\ell^4}}Q^2\bigg].
\end{align}
The polynomial $Y$ has real and distinct roots for $\Delta(Y)>0$, or
\begin{align}
 \brac{\frac{1}{A^2\ell^2}+\Omega_-}Q<E^2<\brac{\frac{1}{A^2\ell^2}+\Omega_+}Q,
\end{align}
where $\Omega_\pm$ is the same as in Eq.~\Eqref{Omega_def}. As $E^2$ approaches the upper bound, $y_1$ approaches $y_2$, whereas if $E^2$ approaches the lower bound, $y_2$ approaches $y_*$. Now, in the AdS ($\ell^2>0$) case the lower bound consists of a positive and negative term, and one may wonder whether this gives a positive non-zero lower bound for $E^2$. However, we can show that this lower bound is always negative as long as $F$ has distinct real roots. Recalling Eq.~\Eqref{Froots}, this condition requires $\ell^2>\frac{54m^2A^2\Omega_+}{2A^2\brac{1-4m^2A^2}^2}$. Upon rearranging, we have 
\begin{align}
 \frac{1}{A^2\ell^2}<\frac{2\brac{1-4m^2A^2}^2}{54m^2A^2\Omega_+}.
\end{align}
Using the relation $\Omega_+\Omega_-=-\frac{\brac{1-4m^2A^2}^2}{27m^2A^2}$, one directly computes 
\begin{align}
 \frac{1}{A^2\ell^2}+\Omega_-<0.
\end{align}
Therefore this lower bound could not be reached by a positive $E^2$.

Next we organise the different cases as we increase $E^2$ continuously:

\textbf{Case I, $0<E^2<\brac{\Omega_++\frac{1}{A^2\ell^2}}Q$.} This is the case depicted in Fig.~\ref{fig_FXgraph}, where $Y$ has distinct roots. A photon may propagate in one of the two possible disconnected domains 
\begin{align}
 a<y\leq y_1\quad\mbox{ or }\quad y_1\leq y<b.
\end{align}
In other words, there is a `potential barrier' at $y_1<y<y_2$ which prevents photons in $a<y\leq y_1$ to access $y_1\leq y<b$, and vice versa. As $E^2$ is increased, the potential barrier narrows and $y_1$ and $y_2$ approach each other. 

\textbf{Case II, $E^2=\brac{\Omega_++\frac{1}{A^2\ell^2}}Q$.} Here $y_1$ and $y_2$ coalesce into a degenerate root given by 
\begin{align}
 y_1=y_2=y_{\mathrm{ps}}=\frac{-1-\sqrt{1+12m^2A^2}}{6mA}.
\end{align}
In this case, $Y(y_{\mathrm{ps}})=Y'(y_{\mathrm{ps}})=0$, and Eq.~\Eqref{ydot} is solved by $y$ being constant at $y_{\mathrm{ps}}$. This is the \emph{aspherical photon surface} of the C-metric. This terminology was introduced by Gibbons and Warnick \cite{Gibbons:2016isj} where they have shown that the geometry of constant-$y$ surfaces is that of a deformed sphere. This is also the analogue to the photon spheres around the Kerr black hole \cite{Teo:2003}.

\textbf{Case III, $E>\brac{\Omega_++\frac{1}{A^2\ell^2}}Q$.} In this case $y_1$ and $y_2$ become complex conjugate pairs. Then $F$ is always positive in the Lorentzian region $a<y<b$ and the potential barrier no longer exists. So photons from the neighbourhood of the black-hole horizon have access to the acceleration horizon, and vice versa.  

\begin{figure}
 \begin{center}
  \includegraphics{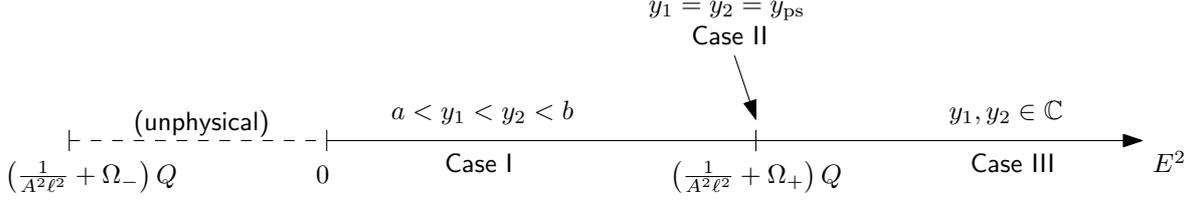}
  \caption{The root structure of $Y$, for various values of $E^2$. The quantity $\Omega_\pm$ is as defined in Eq.~\Eqref{Omega_def}.}
  \label{fig_Yrange}
 \end{center}

\end{figure}

The cases described above are summarised in Fig.~\ref{fig_Yrange}, showing the different cases as $E^2$ varies from $E^2>0$.

\subsection{Organisation of parameter ranges and examples}
 
In a metric of mass parameter $m$, acceleration parameter $A$ and (A)dS parameter $\ell^2$, a photon trajectory can take either of the following cases classified by A, B, C, or D based on its angular momentum, and cases I, II, or III based on its energy. We have seen that the root structure of $X$ depends on $\Phi^2/Q$, whereas the root structure of $Y$ depends on $E^2/Q$. It is then convenient to define dimensionless quantities 
\begin{align}
 \eta=\frac{\Phi^2}{Q},\quad\xi=\frac{E^2}{Q}.
\end{align}
The type of motion can now be parametrised in $(\eta,\xi)$-space subdivided into 12 domains which we denote AI, AII, AIII, BI, \ldots, EIII. For example, a photon trajectory with angular momentum in the range $1<\eta<1$ (Case B) and with energy $\xi=\Omega_++\frac{1}{A^2\ell^2}$ (Case I), will be called Case BII. For another example, the circular photon orbits of constant $x$ and $y$ belongs to EII, where $\brac{\eta,\xi}=\brac{-\Omega_-,\Omega_++\frac{1}{A^2\ell^2}}$.  

\begin{figure}
 \begin{center}
  \includegraphics{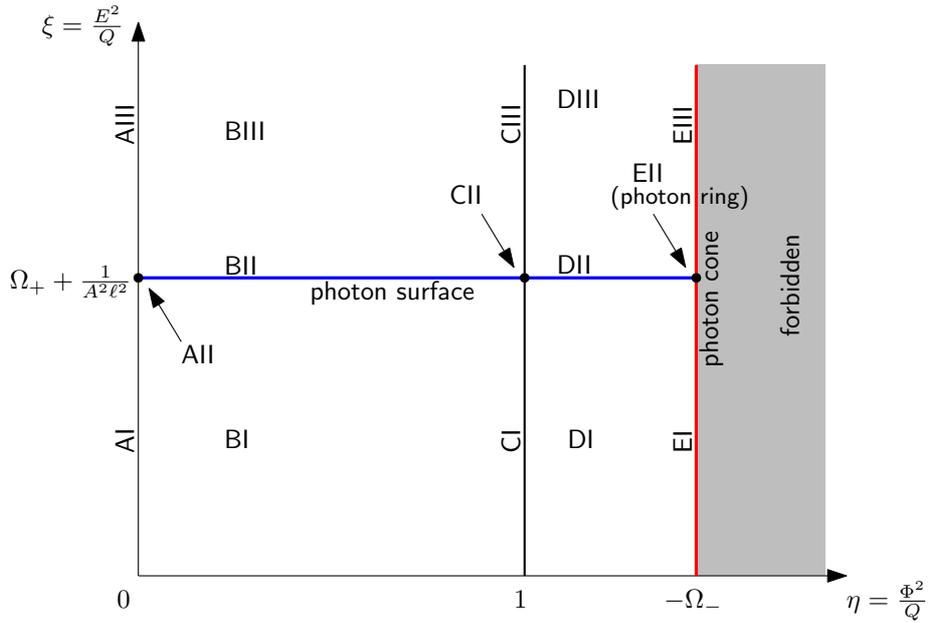}
  \caption{(Colour online) The parameter space of null geodesics in the (A)dS C-metric. The relevant domains are labelled with the Latin alphabet followed by Roman numerals, where the Latin parts `A', `B',\ldots denote the cases in the root structure analysis for $X(x)$, and the Roman numerals `I', `II', and `III' denote the cases listed in the analysis for $Y(y)$. The blue horizontal line denotes the photon surfaces of constant $y$, and the red vertical line denotes the photon cones of constant $x$. The intersection point $(-\Omega_-,\Omega_++\frac{1}{A^2\ell^2})$ corresponds to the parameters required for the circular photon orbits, where $\Omega_\pm$ are given by Eq.~\Eqref{Omega_def}.}
  \label{fig_ParamSpace}
 \end{center}

\end{figure}

The respective domains for each case are sketched in Fig.~\ref{fig_ParamSpace}. The cases AI, AIII, BII, CI, CIII, DII, EI, and EIII occupy one-dimensional lines, while AII, CII, and EII are points. The aspherical photon surfaces are represented by the blue horizontal line, starting from the part where $0\leq\eta<1$ is the case AII, for which the domain of polar oscillation crosses the `equatorial plane' $x=0$. For $1<\eta<-\Omega_-$ on this line is where the surface lies completely in the `northern hemisphere' $x>0$ and never crosses the `equatorial plane'. The critical case between the two is the point BII, where the southernmost part of the polar oscillation just touches the equatorial plane. The vertical red line represents $\ell=-\Omega_-$ where the geodesics have constant $x$, containing the cases DI, DII, and DIII. The photon surface and constant-$x$ lines intersect at the point $(-\Omega_-,\Omega_++\frac{1}{A^2\ell^2})$, corresponding to the circular photon orbit where both $x$ and $y$ are constant.

Let us now demonstrate some examples of the various cases. The trajectories are obtained by either by solving Eq.~\Eqref{ddots} or using the analytical solutions of Sec.~\ref{analytical}, giving parametric equations for $x$ and $y$. We can convert to more familiar Boyer--Lindquist-type coordinates by the transformation 
\begin{align}
 \theta=\arccos x,\quad r=-\frac{1}{Ay}.
\end{align}
In the Ricci-flat case, these coordinates are those used in Refs.~\cite{Lim:2014qra,Alawadi:2020qdz,Frost:2020zcy}. We further transform to Cartesian-like coordinates by the transformation 
\begin{align}
 X_1=r\sin\theta\cos\phi,\quad X_2=r\sin\theta\sin\phi,\quad X_3=r\cos\theta. \label{cartesian}
\end{align}
For the moment, we assume the azimuthal coordinate has periodicity $\phi\in[0,2\pi)$, leaving both the cosmic strut and string present. As mentioned in Sec.~\ref{EOM} and in the literature, we can remove one of them by appropriately redefining the periodicity of $\phi$. 
For concreteness, the following examples in this section take place in an AdS C-metric with parameters 
\begin{align}
 mA=0.2,\quad \frac{m^2}{\ell^2}=50. 
\end{align}
For these values, we have\footnote{The symbol `$\simeq$' indicates that numerical values are being displayed up to five significant figures.} 
\begin{align}
 \Omega_++\frac{1}{A^2\ell^2}\simeq1.1299,\quad -\Omega_-\simeq1.0373. 
\end{align}

\begin{figure}
 \begin{center}
  \includegraphics[width=0.55\textwidth, trim= 40 40 40 40, clip=true]{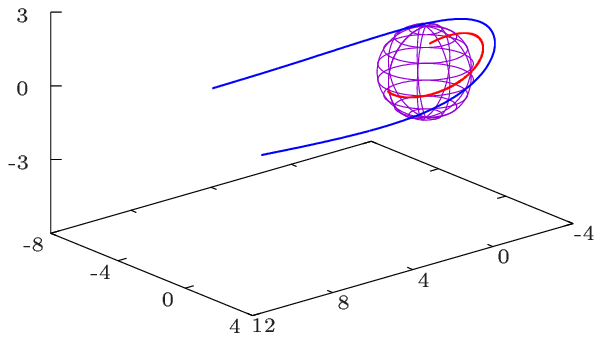}
  \includegraphics[width=0.4\textwidth]{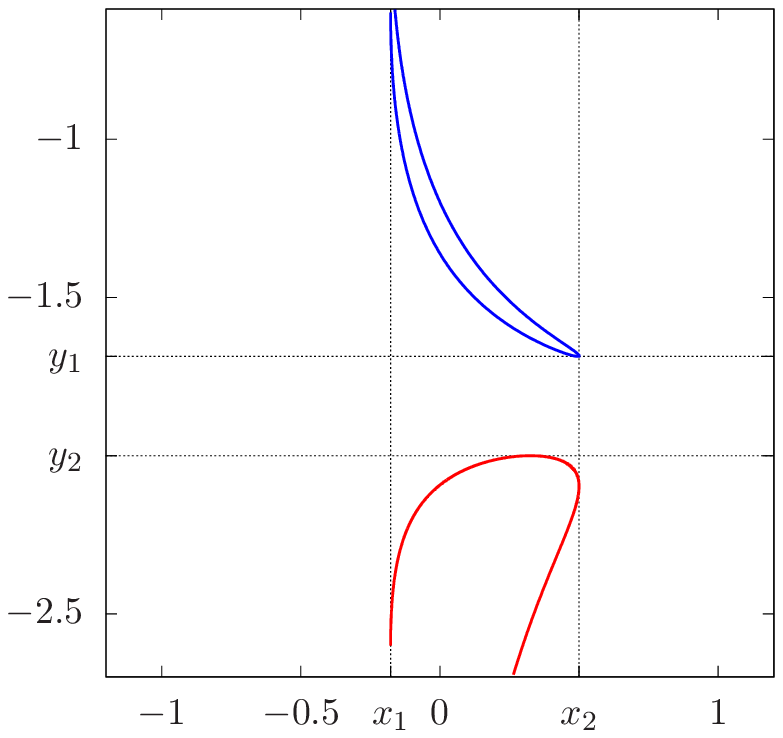}
 \end{center}
 \caption{(Colour online) Examples of orbits from case BI in a spacetime with parameters $mA=0.2$, $\ell^2=\frac{m^2}{50}$. The left panel shows the trajectory in the Cartesian-like coordinates defined in Eq.~\Eqref{cartesian} while the right panel is in $(x,y)$-coordinates. The energy and angular momentum parameters are $\eta=0.9$, $\xi=1.1$. The blue line represents a photon starting just before the acceleration horizon at $y=-0.6$ and the red curve represents a photon starting just outside the black hole horizon at $y=-2.6$ heading outwards.}
 \label{fig_ExBI}

\end{figure}

We start with an example from Case BI shown in Fig.~\ref{fig_ExBI}, with $(\eta,\xi)=(0.9,1.1)$. In this case, geodesics can propagate in two disconnected domains separated by a potential barrier $y_1<y<y_2$. The blue curves in Fig.~\ref{fig_ExBI} represent a photon starting just before the acceleration horizon at $y=-0.6$ towards the black hole, encountering the potential barrier at $y_1$ before going off into the acceleration horizon. The red curve represents a photon starting just outside the black hole horizon at $y=-2.6$ heading outwards. It encounters the potential barrier at $y_2$ and subsequently falls into the black hole.

\begin{figure}
 \begin{center}
  \begin{subfigure}[b]{\textwidth}
    \includegraphics[width=0.55\textwidth, trim= 40 40 40 40, clip=true]{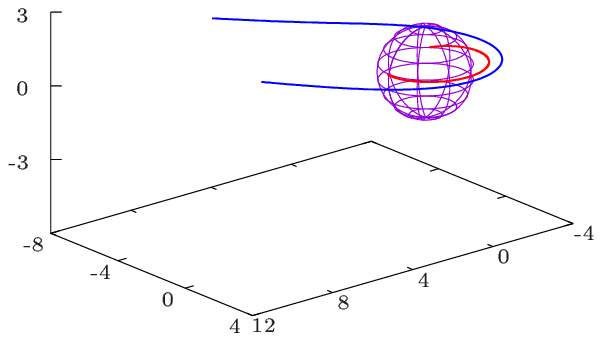}
    \includegraphics[width=0.4\textwidth]{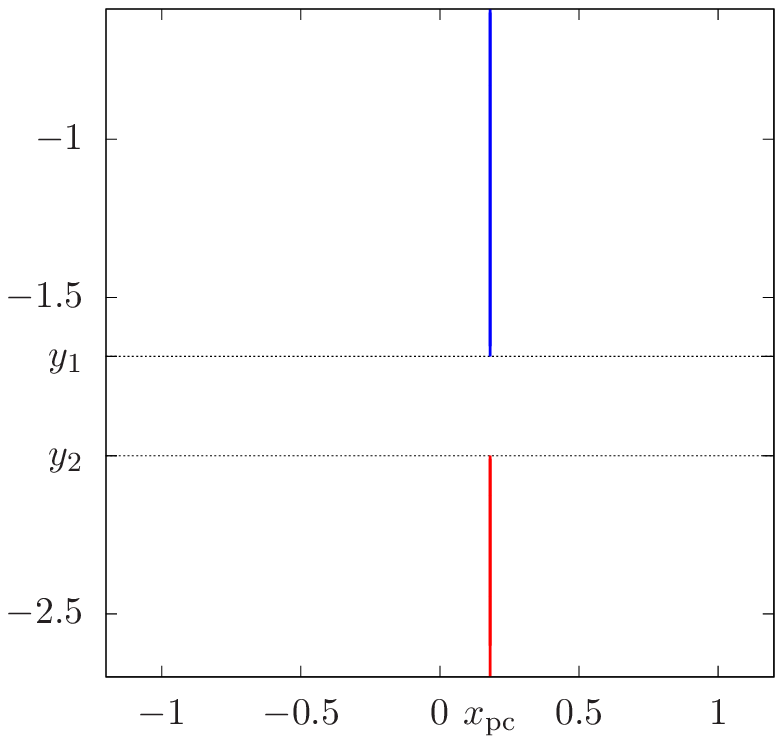}
    \caption{Case EI, with initial position $y=-0.6$ (blue) and $y=-2.6$ (red).}
    \label{fig_ExEI}
  \end{subfigure}
  \begin{subfigure}[b]{\textwidth}
    \includegraphics[width=0.55\textwidth, trim= 50 50 50 50, clip=true]{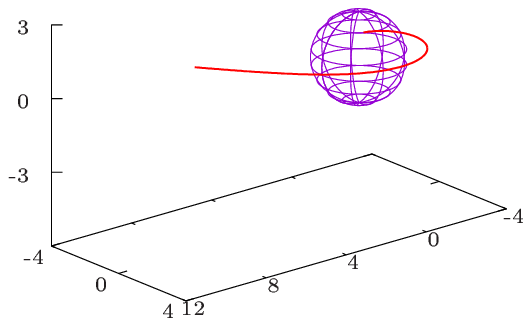}
    \includegraphics[width=0.4\textwidth]{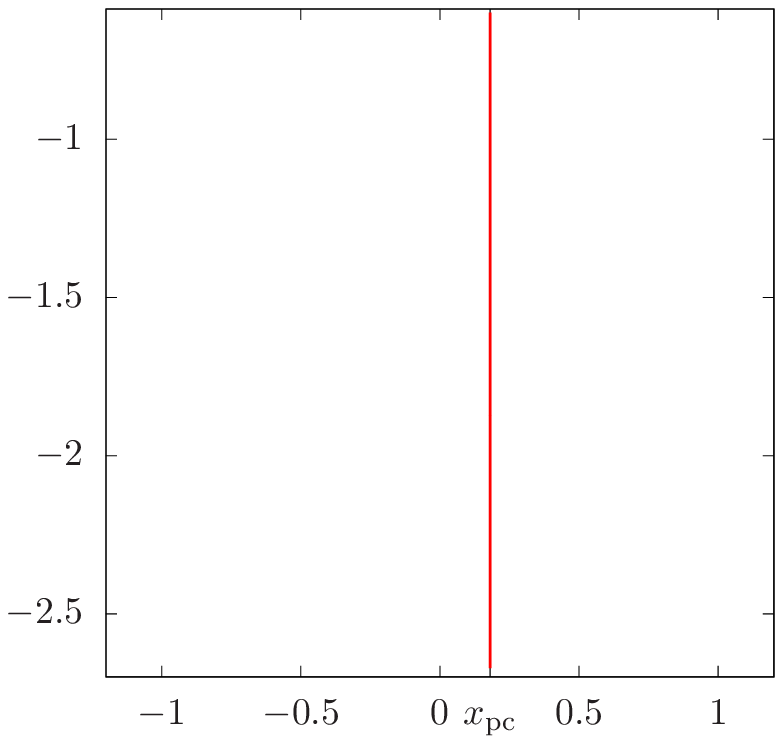}
    \caption{Case EIII, with initial position $y=-0.6$.}
    \label{fig_ExEIII}
  \end{subfigure}
  \end{center}
  \caption{(Colour online) Examples of orbits on the photon cone, with Case EI (a) and Case EIII (b).}
  \label{fig_ExPhotonCone}

\end{figure}

For the next example let us consider photon cones where $\eta=-\Omega_-$. Here the geodesics are confined to $x_{\mathrm{pc}}=\mathrm{constant}$.  In Fig.~\ref{fig_ExPhotonCone} we show explicitly the cases EI and EIII. For the case EI shown in Fig.~\ref{fig_ExEI}, the blue curve shows the photon initially at a position just before the acceleration horizon, $y=-0.6$, heading towards the black hole until it encounters the potential barrier $y_1$, before heading away and falling beyond the acceleration horizon. The red curve is a trajectory starting just outside the black-hole horizon, $y=-2.6$ heading outwards before encountering the potential barrier at $y_2$, upon which it turns back and falls into the black hole. The case EIII example is shown in Fig.~\ref{fig_ExEIII}, showing the photon initially at $y=-0.6$ taking a trajectory that eventually falls into the black-hole horizon. At all times the $x$ coordinate is constant at $x=x_{\mathrm{pc}}$.

\begin{figure}
 \begin{center}
  \begin{subfigure}[b]{\textwidth}
    \includegraphics[width=0.55\textwidth, trim= 20 90 20 0, clip=true]{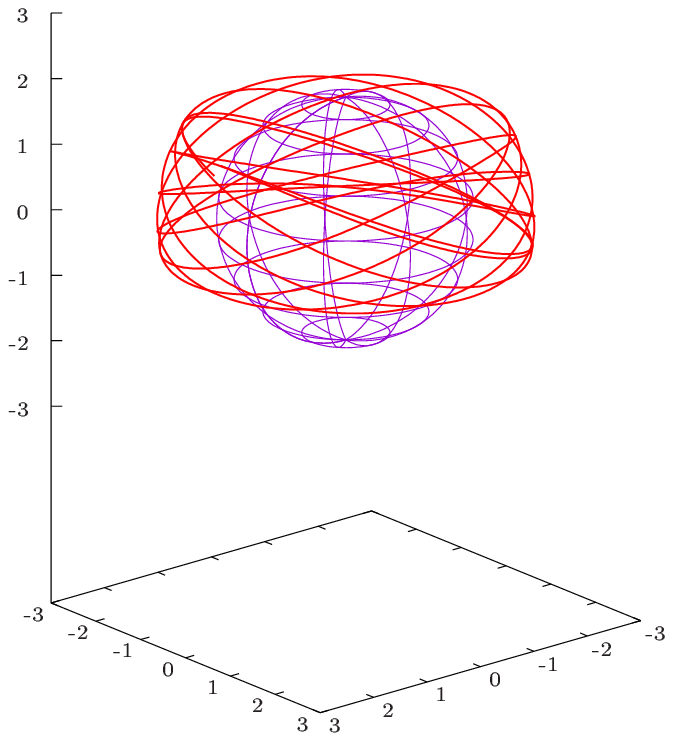}
    \includegraphics[width=0.4\textwidth]{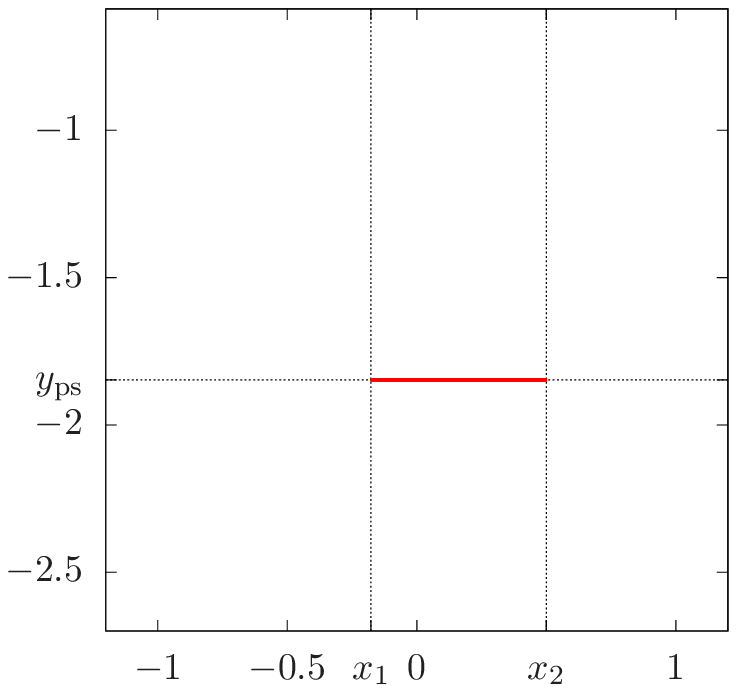}
    \caption{Case BII, with $(\eta,\xi)=(0.9,\,\Omega_++\frac{1}{A^2\ell^2})$.}
    \label{fig_ExBII}
  \end{subfigure}
  \begin{subfigure}[b]{\textwidth}
    \includegraphics[width=0.55\textwidth, trim= 20 90 20 0, clip=true]{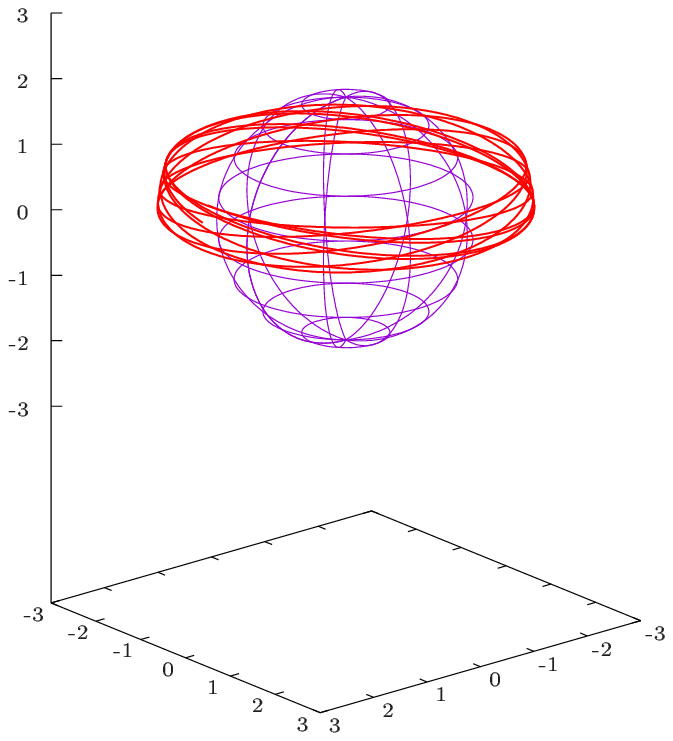}
    \includegraphics[width=0.4\textwidth]{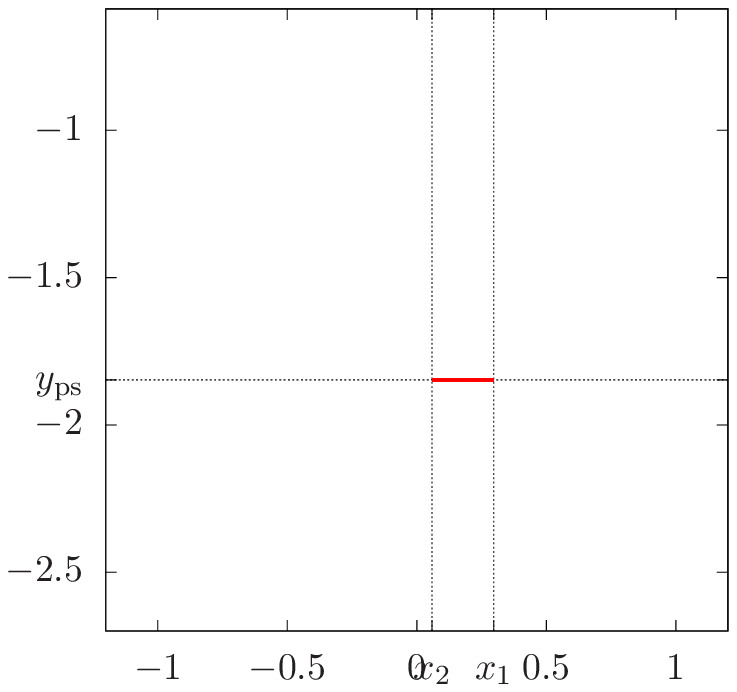}
    \caption{Case DII, with $(\eta,\xi)=(1.02,\,\Omega_++\frac{1}{A^2\ell^2})$.}
    \label{fig_ExDII}
  \end{subfigure}
  \begin{subfigure}[b]{\textwidth}
    \includegraphics[width=0.55\textwidth, trim= 20 90 20 0, clip=true]{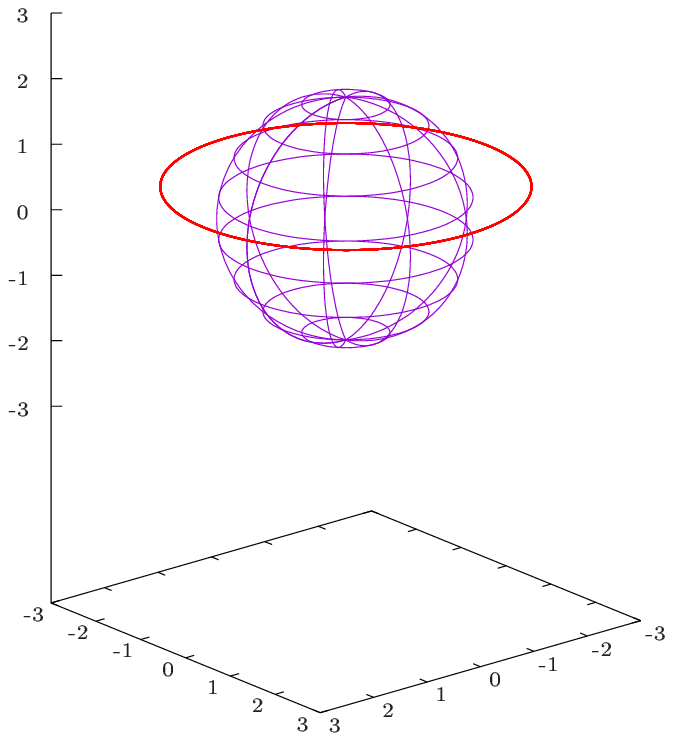}
    \includegraphics[width=0.4\textwidth]{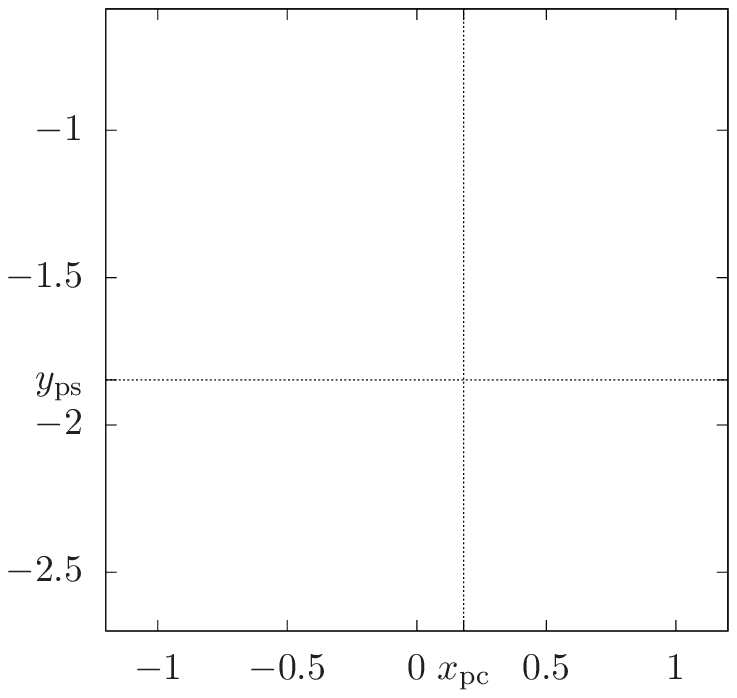}
    \caption{Case EII, with $(\eta,\xi)=(-\Omega_-,\,\Omega_++\frac{1}{A^2\ell^2})$.}
    \label{fig_ExEII}
  \end{subfigure}
  \end{center}
  \caption{(Colour online) Orbits on the photon surface.}
  \label{fig_ExPhotonSurface}

\end{figure}

Perhaps the more interesting case would be the aspherical photon surface with $\xi=\Omega_++\frac{1}{A^2\ell^2}$ and $y$ is constant at $y_{\mathrm{ps}}$. The examples are depicted in Fig.~\ref{fig_ExPhotonSurface}. In Fig.~\ref{fig_ExBII}, we show a geodesic from case BII for $\eta=0.9$. Here the geodesics oscillates in the polar direction that contains the `equatorial plane'. Increasing $\eta$ to $1.02$, the domain of the $x$-oscillation narrows. This value of $\eta$ corresponds to case DII shown in Fig.~\ref{fig_ExDII}, where $x_2$ is positive. So the photon oscillates in a domain contained in the `northern hemisphere' and does not cross the `equatorial plane'. Further increasing $\eta$ to the value $-\Omega_-$, we now have $x$ also being constant at $x_{\mathrm{pc}}$. This is now a circular photon ring of Case EII, shown in Fig.~\ref{fig_ExEII}.

\section{Analytical solutions} \label{analytical}

Analogous to the case of Kerr geodesics, it will be convenient to introduce a Mino-like parameter \cite{Mino:2003yg} $\lambda$, which is defined in the present case by $\frac{\dif\lambda}{\dif\tau}=A^{2}(x-y)^{2}$. Under this reparametrisation Eqs.~\Eqref{tdot}--\Eqref{ydot} appear as
\begin{subequations}
\begin{align}
 \frac{\dif t}{\dif\lambda}&=\frac{E}{F},\label{tdot_M}\\
 \frac{\dif\phi}{\dif\lambda}&=\frac{\Phi}{G},\label{phidot_M}\\
 \frac{\dif x}{\dif\lambda}&=\pm\sqrt{X(x)},\label{xdot_M}\\
 \frac{\dif y}{\dif\lambda}&=\pm\sqrt{Y(y)},\label{ydot_M} 
\end{align}
\end{subequations}
so that the equations for $x$ and $y$ can be integrated directly as functions of $\lambda$.

\subsection{Solution for \texorpdfstring{$x(\lambda)$}{x}}

Here, it is convenient to express $X$ in terms of its roots. Here we shall assume the roots are distinct (corresponding to cases A--D), since there are no physically relevant geodesics for when $X$ has complex roots, and the case with degenerate roots simply corresponds to the photon cones of constant $x$. Therefore we write 
\begin{align}
 X&=2mAQ(x_2-x)(x-x_1)(x-x_*).
\end{align}
Choosing the branch and the initial conditions will determine the specific forms of the solutions. When a geodesic encounters a turning point where $X=0$, care must be taken in correctly choosing the correct branch for its subsequent motion. Essentially one must always choose the branch such that future-directed geodesics remain future-directed, and similarly for the past-directed geodesics. Future-directed geodesics are those where $\lambda$ increases throughout the motion, since by Eq.~\Eqref{tdot_M} lead to $t$ increasing as well, and the opposite holds for past-directed geodesics.

Firstly, suppose we take the initial conditions $x(0)=x_2$. In this case we choose the lower sign of Eq.~\Eqref{xdot_M} so that $\lambda$ increases as the $x$-coordinate of the photon decreases away from $x_2$. Then, Eq.~\Eqref{xdot_M} can be integrated as 
\begin{align}
 \frac{1}{\sqrt{2mAQ}}\int_x^{x_2}\frac{\dif x'}{\sqrt{(x_2-x')(x'-x_1)(x'-x_*)}}=\int_0^{\lambda}\dif\lambda'
\end{align}
This integral can be evaluated exactly (see, e.g., \cite{gradshteyn2014table}). The result is\footnote{Note that $\lambda$ lies outside the square root.} 
\begin{align}
 \mathrm{F}\brac{\arcsin\sqrt{\frac{x_2-x}{x_2-x_1}},\sqrt{\frac{x_2-x_1}{x_2-x_*}}}&=\sqrt{\half mAQ(x_2-x_*)}\;\lambda,
\end{align}
where $\mathrm{F}(\varphi,k)$ is the elliptic integral of the first kind. Inverting to express $x$ as a function of $\lambda$, we get 
\begin{align}
 x(\lambda)=x_2-(x_2-x_1)\mathrm{sn}\brac{\sqrt{\half mAQ(x_2-x_*)}\;\lambda,\; \sqrt{\frac{x_2-x_1}{x_2-x_*}}}^2,
\end{align}
where $\mathrm{sn}(\psi,q)$ is the Jacobi elliptic function of the first kind.

On the other hand, if we take the initial conditions $x(0)=x_1$, we choose the upper sign of Eq.~\Eqref{xdot_M} so that $\lambda$ increases as the $x$-coordinate of the photon increases away from the southern boundary $x_1$. In this case the integral is 
\begin{align*}
 \frac{1}{\sqrt{2mAQ}}\int_{x_1}^x\frac{\dif x'}{(x_2-x')(x'-x_1)(x'-x_*)}&=\int_0^\lambda\dif\lambda'\nonumber\\
 \mathrm{F}\brac{\arcsin\sqrt{\frac{(x_2-x_*)(x-x_1)}{(x_2-x_1)(x-x_*)}},\sqrt{\frac{x_2-x_1}{x_2-x_*}}}&=\sqrt{\half mAQ(x_2-x_*)}\;\lambda.
\end{align*}
Inverting this to solve for $x$, we have 
\begin{align}
 x(\lambda)&=\frac{(x_2-x)x_1-x_*(x_2-x_1)\mathrm{sn}\brac{\sqrt{\half mAQ(x_2-x_*)}\;\lambda\;,\sqrt{\frac{x_2-x_1}{x_2-x_*}}}^2}{x_2-x_*-(x_2-x_1)\mathrm{sn}\brac{\sqrt{\half mAQ(x_2-x_*)}\;\lambda\;,\sqrt{\frac{x_2-x_1}{x_2-x_*}}}^2}.
\end{align}

\subsection{Solution for \texorpdfstring{$y(\lambda)$}{y}}

Similar to the solutions for $x(\lambda)$ the specific forms of the solutions depend on the initial conditions and the choice of branch. As before we shall choose the branches that correspond to future-directed geodesics. Here we shall consider Cases I and III, since Case II is simply the photon surface where $y$ is constant. 

We first consider Case I, where the roots of $Y$ are real and distinct. Hence we write 
\begin{align}
 Y=2mAQ(y_*-y)(y-y_2)(y-y_1),\quad y_1<y_2<y_*.
\end{align}
In this case the geodesics can either exist in $y>y_2$ or $y<y_1$. In the former case, we take the initial conditions $y(0)=y_2$. We choose the upper branch of the square root so that $\lambda$ increases as the photon increases in $y$ away from $y_2$ (and heads towards the acceleration horizon). Then Eq.~\Eqref{ydot_M} can be integrated as
\begin{align}
 \frac{1}{\sqrt{2mAQ}}\int_{y_2}^y\frac{\dif y'}{(y_*-y')(y'-y_2)(y'-y_1)}&=\int_0^\lambda\dif\lambda'\nonumber\\
 \mathrm{F}\brac{\arcsin\sqrt{\frac{(y_*-y_1)(y-y_2)}{(y_*-y_2)(y-y_1)}},\sqrt{\frac{y_*-y_2}{y_*-y_1}}}&=\sqrt{\half mAQ(y_*-y_1)}\;\lambda. \nonumber
\end{align}
We invert this to solve for $y$,
\begin{align}
 y(\lambda)=\frac{(y_*-y_1)y_2-y_1(y_*-y_2)\mathrm{sn}\brac{\sqrt{\half mAQ(y_*-y_1)}\;\lambda,\;\sqrt{\frac{y_*-y_2}{y_*-y_1}}}^2}{y_*-y_1-(y_*-y_1)\mathrm{sn}\brac{\sqrt{\half mAQ(y_*-y_1)}\;\lambda,\;\sqrt{\frac{y_*-y_2}{y_*-y_1}}}^2}.
\end{align}

On the other hand, for photons in the domain $y<y_1$, we shall take $y(0)=y_1$ and choose the lower branch of the square root so that $\lambda$ increases as the photon's $y$-coordinate decreases away from $y_1$ (and heads towards the black hole). Here the integral is 
\begin{align}
 \frac{1}{\sqrt{2mAQ}}\int_{y}^{y_1}\frac{\dif y'}{(y_1-y')(y_2-y')(y_*-y')}&=\int_0^\lambda\dif\lambda'\nonumber\\
 \mathrm{F}\brac{\arcsin\sqrt{\frac{y_1-y}{y_2-y}},\sqrt{\frac{y_*-y_2}{y_*-y_1}}}&=\sqrt{\half mAQ(y_*-y_1)}\;\lambda. \nonumber
\end{align}
Solving for $y$, we obtain 
\begin{align}
 y(\lambda)&=\frac{y_2\mathrm{sn}\brac{\sqrt{\half mAQ(y_*-y_1)}\;\lambda,\sqrt{\frac{y_*-y_2}{y_*-y_1}}}^2-y_1}{\mathrm{sn}\brac{\sqrt{\half mAQ(y_*-y_1)}\;\lambda,\sqrt{\frac{y_*-y_2}{y_*-y_1}}}^2-1}.
\end{align}

Turning to Case III, we now have the situation where $y_1$ and $y_2$ are complex conjugate pairs. We write $y_1=\alpha+\im\beta$, $y_2=\alpha-\im\beta$ for $\alpha,\beta\in\mathbb{R}$. Then, the function $Y$ takes the form 
\begin{align}
 Y=2mAQ(y_*-y)\sbrac{(y-\alpha)^2-\beta^2}.
\end{align}
Let us choose the initial conditions $y(0)=y_0$ for some arbitrary $y_0$. Choosing the lower branch means we have $\lambda$ increasing as $y$ decreases away from its initial position. The integral is 
\begin{align}
 \frac{1}{\sqrt{2mAQ}}\int_{y_0}^y\frac{\dif y'}{\sqrt{(y_*-y')\sbrac{{(y-\alpha)^2-\beta^2}}}}&=-\int_0^\lambda\dif\lambda'\nonumber\\
 \sbrac{\mathrm{F}\brac{\arcsin\sqrt{\frac{y_*-y'}{y_*-\alpha-\beta}},\sqrt{\frac{y_*-\alpha-\beta}{y_*-\alpha+\beta}}  }}_{y_0}^y &=\sqrt{\half mAQ(y_*-\alpha+\beta)}\;\lambda\nonumber
\end{align}
Solving this for $y$, 
\begin{align}
 y(\lambda)&=y_*+(\alpha+\beta-y_*)\mathrm{sn}\brac{\sqrt{\half mAQ(y_*-\alpha+\beta)}(\lambda+C),\;\sqrt{\frac{y_*-\alpha-\beta}{y_*-\alpha+\beta}}}^2,
\end{align}
where $C$ is an integration constant chosen so that $y(0)=y_0$.

\subsection{Equation between \texorpdfstring{$x$}{x} and \texorpdfstring{$\phi$}{phi}}

Here we can obtain an exact expression describing how $\phi$ evolves with $x$. This would be helpful in comparing the frequencies of $x$-oscillations with the angular motion. In particular, it enables us to find periodic or closed orbits, which is where the ratio of the two frequencies are rational numbers. 

Equations.~\Eqref{phidot_M} and \Eqref{xdot_M} together give
\begin{align}
 \frac{\dif\phi}{\dif x}=\pm\frac{\Phi}{G\sqrt{X}}.
\end{align}
Again, the specific form of the solution depends on the choice of initial condition and branch. 

We first consider the initial condition $\phi(x_2)=0$, and choose the lower sign. We have 
\begin{align}
 \int_0^\phi\dif\phi'&=-\frac{1}{\sqrt{2mAQ}}\int_{x_2}^x\frac{\Phi\,\dif x'}{G\sqrt{(x_2-x')(x'-x_1)(x'-x_*)}}. \label{phiint}
\end{align}
To perform this integral, it helps to do a partial fraction decomposition on $1/G$. This turns the right-hand side of \Eqref{phiint} into a sum of three integrals,\footnote{Note that $\phi$ lies outside the square root.}
\begin{align}
 \sqrt{2mAQ}\;\phi&=I_1+I_2+I_3,
\end{align}
where 
\begin{subequations}
\begin{align}
 I_1&=-\frac{\Phi}{2(1+2mA)}\int_{x_2}^x\frac{\dif x'}{(1-x')\sqrt{(x_2-x')(x'-x_1)(x'-x_*)}},\\
 I_2&=-\frac{\Phi}{2(1-2mA)}\int_{x_2}^x\frac{\dif x'}{(1+x')\sqrt{(x_2-x')(x'-x_1)(x'-x_*)}},\\
 I_3&=\frac{4m^2A^2\Phi}{1-4m^2A^2}\int_{x_2}^x\frac{\dif x'}{(1+2mAx')\sqrt{(x_2-x')(x'-x_1)(x'-x_*)}}.
\end{align}
\end{subequations}
Evaluating the integrals, the result is 
\begin{align}
 \phi_2(x)&=\frac{\Phi}{\sqrt{2mAQ(x_2-x_*)}}\bigg[-\frac{1}{(1+2mA)(x_2-1)}\Pi\brac{\zeta,\frac{x_2-x_1}{x_2-1},p}\nonumber\\
   &\hspace{2cm}+\frac{1}{(1-2mA)(x_2+1)}\Pi\brac{\zeta,\frac{x_2-x_1}{x_2+1},p}\nonumber\\
   &\hspace{2cm} -\frac{4mA}{(1-4m^2A^2)(x_2+\frac{1}{2mA})}\Pi\brac{\zeta,\frac{x_2-x_1}{x_2+\frac{1}{2mA}},p}\bigg], \label{phi2}
\end{align}
where $\Pi(\varphi,n,k)$ is the elliptic integral of the third kind, and  
\begin{align}
 \zeta=\arcsin\sqrt{\frac{x_2-x}{x_2-x_1}},\quad p=\sqrt{\frac{x_2-x_1}{x_2-x_*}}. \label{abbrev}
\end{align}

On the other hand, if we consider the initial condition $\phi(x_1)=\phi_0$ for some $\phi_0$, we choose the upper sign for Eq.~\Eqref{xdot_M}. Performing the same partial fraction decomposition, the integral results in 
\begin{align}
 \phi_1(x)&=\phi_0+\frac{\Phi}{\sqrt{2mAQ(x_2-x_*)}}\nonumber\\
  &\,\times\Bigg\{\frac{-1}{(1+2mA)(x_*-1)(x_2-1)}\sbrac{(x_*-x_1)\Pi\brac{\kappa,\frac{x_*-1}{x_1-1}p^2,p}+(x_1-1)\mathrm{F}(\kappa,p)}\nonumber\\
  &\hspace{0.45cm}+\frac{1}{(1-2mA)(x_*+1)(x_2+1)}\sbrac{(x_*-x_1)\Pi\brac{\kappa,\frac{x_*+1}{x_1+1}p^2,p}+(x_1+1)\mathrm{F}(\kappa,p)}\nonumber\\
  &\hspace{1.1cm}-\frac{4mA}{(1-4m^2A^2)(x_*+\frac{1}{2mA})(x_1+\frac{1}{2mA})}\bigg[(x_*-x_1)\Pi\brac{\kappa,\frac{x_*+\frac{1}{2mA}}{x_1+\frac{1}{2mA}}p^2,p}\nonumber\\
     &\hspace{9cm}+\brac{x_1+\frac{1}{2mA}}\mathrm{F}(\kappa,p)\bigg]\Bigg\}, \label{phi1}
\end{align}
where 
\begin{align}
 \kappa=\arcsin\sqrt{\frac{(x_2-x_*)(x-x_1)}{(x_2-x_1)(x-x_*)}}
\end{align}
and $p$ is as defined in Eq.~\Eqref{abbrev}.

\section{Periodic orbits on the photon surface} \label{closed} 

Periodic orbits are closed trajectories in which photons return exactly to their initial condition within a finite $\tau$. Having the analytical solutions of Sec.~\ref{analytical} is particularly useful in seeking out these orbits. In the spirit of Ref.~\cite{Levin:2008mq}, periodic orbits may help in understanding the structure of bound orbits. A simple example of a periodic orbit in the (A)dS C-metric would be the circular photon rings. Besides those, one could also have periodic orbits on the photon surface.

\begin{figure}
 \begin{center}
 \begin{subfigure}[b]{0.49\textwidth}
    \includegraphics[width=\textwidth, trim= 20 90 20 0, clip=true]{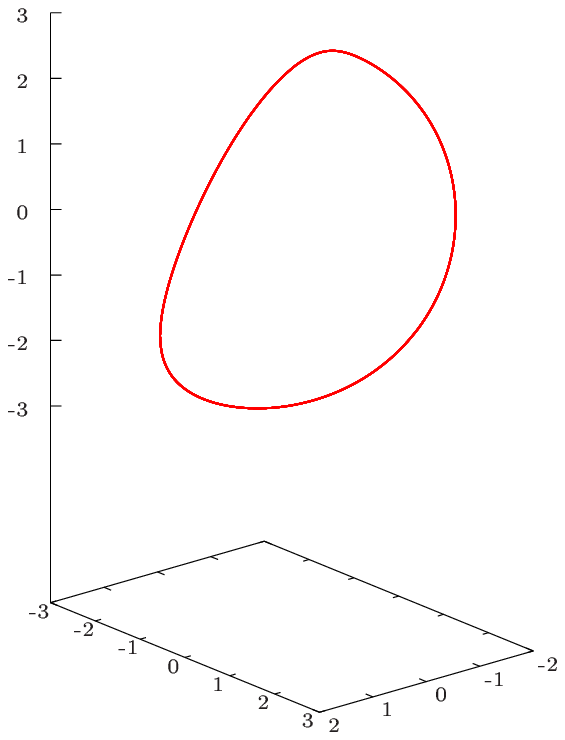}
    \caption{$q=1$, AdS.}
    \label{fig_periodic-AdS}
  \end{subfigure}
  \begin{subfigure}[b]{0.49\textwidth}
    \includegraphics[width=\textwidth, trim= 20 90 20 0, clip=true]{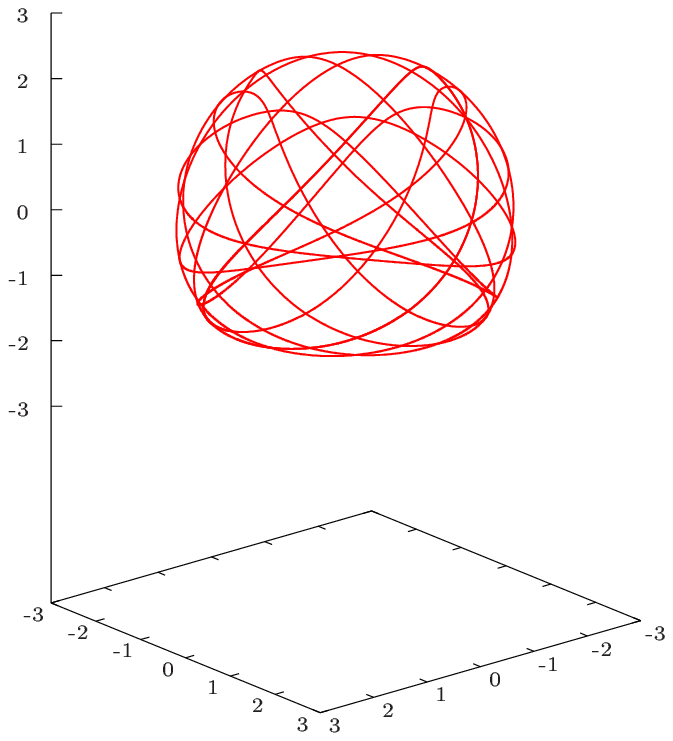}
    \caption{$q=\frac{9}{10}$, dS.}
    \label{fig_periodic-dS}
  \end{subfigure}
  \begin{subfigure}[b]{0.49\textwidth}
    \centering
    \includegraphics[width=\textwidth, trim= 20 90 20 0, clip=true]{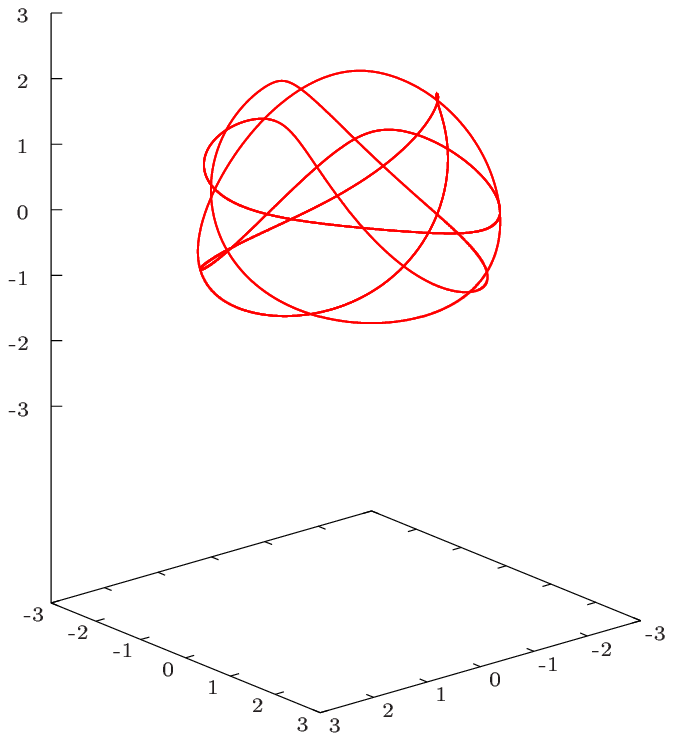}
    \caption{$q=\frac{4}{5}$, Ricci-flat.}
    \label{fig_periodic-flat}
  \end{subfigure}
  \end{center}
  \caption{Examples of periodic photon orbits in the AdS, dS, and Ricci-flat C-metrics. The parameters used in each figure are $mA=0.2$, $\frac{m^2}{\ell^2}=50$,  $\eta=0.25936$ for (a),  $mA=0.3$, $\frac{m^2}{\ell^2}=-0.01$, and $\eta=0.460579$ for (b), and $mA=0.4$, $\frac{m^2}{\ell^2}\rightarrow0$, and $\eta=0.572579$ for (c). To avoid cluttering the figure, we do not plot the black hole horizons.}
  \label{fig_periodic}
\end{figure}

On the photon surface, as the $\phi$-coordinate of the photons revolve around the axis, it also executes oscillations in the $x$-direction. A periodic orbit occurs if, during the interval of one period of $x$-oscillation is completed, the $\phi$-coordinate has changed by a rational multiple of $\Delta\phi$. Therefore, using the analytical solutions of the previous section, we seek 
\begin{align}
 q\Delta\phi&=\phi_1(x_2)+\phi_2(x_1),\quad q\in\mathbb{Q}, \label{period_find}
\end{align}
where $\phi_1(x)$ and $\phi_2(x)$ are as given in Eqs.~\Eqref{phi1} and \Eqref{phi2}, respectively. To find a periodic orbit in a spacetime of a given $m$, $A$, and $\ell$, we fix $\xi=\Omega_++\frac{1}{A^2\ell^2}$ to consider the photon surfaces. For a choice of $\Phi^2$, one can compute the roots $\{x_*,x_1,x_2\}$ and evaluate the right-hand side of \Eqref{period_find}. We then tune $\Phi^2$ until a value corresponding to rational $q$ is found (or at least sufficiently close to desired numerical accuracy).  In Fig.~\ref{fig_periodic}, we plot some examples of periodic orbits found in this way.

\section{Conclusion} \label{conclusion}

In this paper we have derived the equations of null geodesics in the C-metric by means of separating the Hamilton--Jacobi equations. This involves a Carter-like quantity $Q$ that is conserved throughout the motion. Scaling energy and angular momentum in units of $Q$, we obtained a two-dimensional parameter space which characterises the possible types of photon orbits in the C metric. Exact solutions in the $x$ and $y$ coordinates can be obtained with the aid of elliptic integrals.

In the analysis of the coordinate domains, it is interesting to note that there is an obvious asymmetry in the existence domain in the $x$-coordinate. Regarding $x=0$ as the `equatorial plane', the domain of existence is skewed towards the northern hemisphere ($x>0$). This is in contrast to photon spheres around the Kerr black hole where its polar motion is symmetric about the equator. While it is important to remember that $x$ is merely a coordinate and not a physical invariant, this observation is consistent with the intuition that the photons are being `dragged along' by the gravitational attraction of the black hole, which is accelerating towards the southern direction.

\section*{Acknowledgments}
This work is supported by Xiamen University Malaysia Research Fund (Grant No. \\XMUMRF/2019-C3/IMAT/0007).

\bibliographystyle{cmnull}

\bibliography{cmnull}

\end{document}